\documentstyle[12pt]{article}
\begin{document}
\begin{flushright}
DPNU-97-33\\
hep-th/9707141\\
\end{flushright}
\begin{center}
{\Large {\bf Zero Mode and Symmetry Breaking\\ 
 on the Light Front }}\footnote{
To appear in {\it Proc. of International Workshop ``New
Nonperturbative Methods and Quantization on the Light Cone'', Les Houches,
France, Feb. 24 -March 7, 1997.}
}  
\vspace{25pt}
\noindent

Koichi Yamawaki $\mathop{}$\footnote{
E-mail address: yamawaki@eken.phys.nagoya-u.ac.jp}

$\mathop{}$ Department of Physics,  Nagoya University, 
Nagoya 464-01, Japan
\end{center}
\vspace{15pt}
\begin{abstract} 
We discuss the spontaneous symmetry breaking (SSB) on the
light front (LF) in view of the zero mode. 
We first demonstrate impossibility to remove the zero mode in the
continuum LF theory by two examples: The Lorentz invariance forbids even a 
free theory on the LF and the trivial LF vacuum is lost 
in the SSB phase, both due to the zero mode as the accumulating 
point causing uncontrollable infrared singularity. 
We then adopt the Discretized Light-Cone Quantization (DLCQ) which was first 
introduced by Maskawa and Yamawaki to establish the trivial LF
vacuum and 
was re-discovered by Pauli and Brodsky in a different  context.
It is shown in DLCQ that the SSB phase can be 
realized on the trivial LF vacuum only when an 
explicit symmetry-breaking mass of the Nambu-Goldstone (NG) boson $m_{\pi}$  
is introduced as an infrared regulator. 
The NG-boson zero mode integrated over the LF 
must exhibit singular behavior $ \sim 1/m_{\pi}^2$
in the symmetric limit $m_{\pi}\rightarrow 0$ in such a way that
the LF charge is not conserved even in the symmetric limit; $\dot{Q}\ne 0$.
There exists no NG theorem on the LF. Instead,
this singular behavior establishes existence of the massless 
NG boson coupled to
the current whose charge satisfies $Q\vert 0\rangle=0$ and $\dot{Q}\ne 0$, 
in much the same as the NG theorem in the equal-time 
quantization which
ensures existence of the massless NG boson coupled to the current 
whose charge satisfies $Q\vert 0\rangle \ne 0$ and $\dot{Q} =0$.
We demonstrate such a peculiarity in a concrete model, the linear sigma model,
where the role of zero-mode constraint is clarified. 
\end{abstract}
\newpage
\setlength{\baselineskip}{0.3in}

\section{INTRODUCTION}

Much attention has recently been paid to the light-front (LF) 
quantization \cite{Dira} as a promising approach to solve the
nonperturbative dynamics. The most important aspect of the LF quantization
is that the physical LF vacuum is simple, or even trivial \cite{LKS}.
However, such a trivial vacuum, which is vital to the whole LF approach, 
can be realized only if we can 
remove the so-called zero mode out of the physical Fock space 
(``zero mode problem'' \cite{MY}).
  
Actually, the Discretized Light-Cone Quantization (DLCQ) was first introduced 
by Maskawa and Yamawaki \cite{MY}  in 1976 
to resolve the zero mode
problem and was re-discovered by Pauli and
Brodsky \cite{PB} in 1985 in a different context.
The zero mode in DLCQ is clearly isolated from other modes and hence
can be treated in a well-defined manner without ambiguity, 
in sharp contrast to
the continuum theory where the zero mode is the accumulating point and hard to
be controlled in isolation \cite{NY}.  
In DLCQ, Maskawa and Yamawaki \cite{MY} in fact discovered a
constraint equation for the zero mode (``{\it zero-mode constraint}'')
 through which 
the zero mode becomes dependent on other
modes and then they observed that the zero mode can be removed from 
the physical
Fock space by solving the zero-mode constraint, thus {\it establishing the 
trivial LF vacuum in DLCQ}. 
 
Such a trivial vacuum, on the other hand, might confront the usual picture of
complicated nonperturbative vacuum structure 
in the equal-time quantization corresponding to   
confinement, spontaneous 
symmetry breaking (SSB), etc.. Since the vacuum is proved trivial in 
DLCQ \cite{MY},
the only possibility to realize such phenomena would be through
the complicated structure of the operator  
and only such an operator would be the zero mode.
In fact the zero-mode constraint itself implies that 
the zero mode carries essential information on the complicated dynamics.
One might thus expect that explicit solution of the zero-mode constraint
in DLCQ would give rise to the SSB 
while preserving the trivial LF vacuum.
Actually, several authors have recently argued in (1+1)
dimensional models that the solution to the zero-mode constraint
might induce spontaneous
breaking of the discrete symmetries \cite{Rob}. However,
the most outstanding feature of the SSB is
the existence of the Nambu-Goldstone (NG) boson for the continuous symmetry
breaking in (3+1) dimensions. 
 
In this talk I shall explain, within the canonical DLCQ 
\cite{MY}, how  the NG 
phenomenon is realized through the zero modes in (3+1) dimensions
while keeping the vacuum trivial. The talk is based on recent works done in 
collaboration with Yoonbai Kim
and Sho Tsujimaru \cite{KTY,TY}. 
We encounter a striking feature of the
zero mode of the NG boson: Naive use of the zero-mode constraints
does not lead to the NG phase at all ((false) ``no-go theorem'') in contrast
to the current expectation mentioned above. 
It is inevitable to introduce an infrared regularization by
explicit-breaking mass of the NG boson $m_\pi$. 
The NG phase can only be realized via peculiar
behavior of the zero mode of the NG-boson fields:
{\it The NG-boson zero mode, when integrated over the LF, 
must have a singular behavior $\sim 1/m_{\pi}^2$} in the symmetric limit
$m_\pi^2 \rightarrow 0$.
This we demonstrate both in
a general framework of the LSZ reduction formula  
and in a concrete field theoretical model, 
the linear $\sigma$ model.
The NG phase is in fact realized in such a way that the 
{\it vacuum is trivial}
while {\it the LF charge is not conserved} in the symmetric limit 
$m_\pi^2 \rightarrow 0$.

\section{ZERO MODE PROBLEM IN THE CONTINUUM THEORY}

Before discussing the zero mode in DLCQ, we here comment that
{\it it is impossible to remove the zero mode in the continuum theory} 
in a manner consistent with the trivial vacuum, {\it as
far as we use the canonical commutator} \footnote{
We choose the LF ``time'' as $x^{+}=x_{-}\equiv \frac{1}{\sqrt{2}}(x^{0}
+x^{3})$ and the longitudinal and the transverse coordinates are
denoted by $\vec{x} \equiv (x^{-},x^{\bot})$, with $x^{-}\equiv
\frac{1}{\sqrt{2}}(x^{0}-x^{3})$ and $x^{\bot}\equiv
(x^{1},x^{2})$, respectively. 
}
:
\begin{equation}
[\phi(x), \phi(y)]_{x^+=y^+}=-\frac{i}{4}\epsilon(x^--y^-)
\delta^{(2)} (x^{\bot}-y^{\bot}) ,
\label{FSf}
\end{equation}
where the sign function $\epsilon (x^-)$ is defined by
\begin{equation}
\epsilon (x^-) =\frac{i}{\pi} {\cal P} \int_{-\infty}^{+\infty}
\frac{dp^+}{p^+} e^{-ip^+x^-}
\end{equation}
through the principal value prescription 
and hence has no $p^+\equiv 0$ mode.
The point is that the real problem with the zero mode in the
continuum theory is {\it not a single mode} 
with $p^+ \equiv 0$, which is just measure zero, 
but actually the {\it accumulating point} 
$p^+ \rightarrow 0$ \cite{NY}.

To demonstrate this, let us first illustrate a no-go theorem found by 
Nakanishi and Yamawaki \cite{NY}.
The LF canonical commutator (\ref{FSf}) gives explicit expression of Wightman
 two-point
function on LF: 
\begin{equation}
\left. \langle 0|\phi(x)\phi(0)|0\rangle \right|_{x^+=0}
= \frac{1}{2\pi}\int_{0}^{\infty}\frac{dp^+}{2p^+}e^{-ip^+x^-}\cdot 
\delta^{(2)}(x^{\bot}) ,
\label{WFLF}
\end{equation}
which is logarithmically divergent at $p^+ = 0$ and local
in $x^{\bot}$ and, more importantly, is
independent of the interaction and the mass. 
It is easy to see that this is wrong, for example, in the free theory where
the Lorentz-invariant two-point Wightman function is given at any point $x$ by
\begin{equation}
\Delta^{(+)}(x) =\frac{1}{(2\pi)^3}\int_{0}^{\infty}
\frac{dp^+}{2p^+}e^{-ip^-x^+ -ip^+x^- +ip^{\bot}x^{\bot}}
=\frac{m}{4\pi^2\sqrt{-x^2}}K_1 (m\sqrt{-x^2}),
\label{WFcov}
\end{equation}
where $K_1$ is the Hankel function. 
Restricting (\ref{WFcov}) to the LF, $x^+=0$, yields 
$\frac{m}{4\pi^2\sqrt{x_{\bot}^2}}K_1 (m\sqrt{x_{\bot}^2})$,
which is finite (positive definite), nonlocal in $x^{\bot}$ and
dependent on mass, in obvious contradiction to the above result (\ref{WFLF}).
Actually, the Lorentz-invariant result (\ref{WFcov}) is 
a consequence of the {\it mass-dependent} regularization of $1/p^+$ 
singularity by the infinitely oscillating (mass-dependent) 
phase factor $e^{-ip^-x^+}=e^{-i(m^2+p_{\bot}^2)/2p^+ \cdot x^+}$
{\it before taking the LF restriction}
 $x^+ = 0$. Namely, there exists {\it no free theory on the LF}!
This difficulty also applies to the interacting theory satisfying the Wightman 
axioms \cite{NY}.
Thus
the LF restriction from the beginning loses all the information of dynamics 
carried  by the {\it zero mode as the accumulating point}.

Next we show  another difficulty of the continuum LF theory,
that is,  {\it as far as the sign function is
used}, the {\it LF charge does not annihilate the vacuum} in the SSB phase
and hence the trivial LF vacuum is never realized \cite{TY}.   
Let me illustrate this in the $O(2)$ $\sigma$ model where the fields
$\sigma,\pi$ obey the canonical commutator (\ref{FSf}), which yields
\begin{equation}
[Q, \sigma^{\prime}(x)]=-i\pi(x), \quad
[Q, \pi(x)]=i\sigma^{\prime}(x)+\frac{i}{2}v , 
\label{FSi1}
\end{equation}
where the  LF charge is defined as usual by 
$
Q=\int d^3 \vec{x}( \pi\partial_{-}\sigma^{\prime}-
\sigma^{\prime}\partial_{-}\pi
-v\partial_{-}\pi) , 
$
and the anti-periodic boundary condition (B.C.)
 is imposed on $\pi$ and the
shifted field $\sigma^{\prime}\equiv \sigma-v$ ($v=\langle \sigma \rangle$) 
in order to eliminate the surface terms in (\ref{FSi1}), 
 $\phi(x^-=\infty)+\phi(x^-=-\infty)=0$.
The non-zero constant term on the R.H.S. of (\ref{FSi1}) has its 
origin in the commutation relation (\ref{FSf}), which is 
consistent with the anti-periodic B.C.,
$\pi(x^-=\infty)=-\pi(x^-=-\infty)\ne 0$ \footnote{
Without specifying B.C., we would not be able to
formulate consistently the LF quantization \cite{STH,TY}.
If we used anti-periodic B.C. for the {\it full field}
$\sigma$, we would have no zero mode (no $v$) and hence no symmetry 
breaking anyway.
On the other hand, vanishing fields at $x^{-}=\pm\infty$
contradict the commutation relation (\ref{FSf}).
}
.
Then we have
\begin{equation}
\langle 0 \vert [Q, \pi(x)]\vert 0 \rangle
=i\langle 0 \vert 
\sigma^{\prime}(x)\vert 0 \rangle+\frac{i}{2}v  
=\frac{i}{2}v\ne 0 . 
\label{nonannihilation}
\end{equation}
Namely, the {\it LF charge does not annihilate the vacuum},
$Q\vert 0\rangle \ne 0$, due to the zero mode
as the accumulating point,
even though we have ``removed'' exact zero mode $p^+\equiv 0$ by shifting
 the field $\sigma$
to  $\sigma^{\prime}$. This implies that there in fact {\it exists a zero-mode
state} $\vert \alpha\rangle \equiv Q\vert 0 \rangle$ with zero eigenvalue 
of $P^+$ such that 
$P^+\vert \alpha\rangle = [P^+,Q]\vert0\rangle =0$ (due to $P^+$ conservation). 
Our result disagrees with Wilson et al. \cite{Wils} who 
claim to have eliminated the zero mode (to be compensated by  
``unusual counter terms'') in the continuum LF theory.

\section{NAMBU-GOLDSTONE BOSON ON THE LIGHT FRONT}

In contrast to the continuum LF theory mentioned above, 
we already mentioned in Introduction that 
the trivial vacuum is always realized in DLCQ \cite{MY}.
We now discuss how such a trivial vacuum can be reconciled with the 
SSB phenomena. Here we use DLCQ \cite{MY,PB},
$x^{-}\in [-L,L]$,  with a periodic boundary 
condition in the $x^{-}$ direction, and then
take the continuum limit $L \rightarrow \infty$
in the end of whole calculation. 
We should mention here that the above no-go theorem \cite{NY} cannot be solved  
by simply taking the continuum limit of the DLCQ 
nor by any other existing method and would be a future problem to 
be solved in a more profound way.

\subsection{``NO-GO THEOREM'' (FALSE)}

Let us first prove a ``no-go theorem'' (which will turn out to be false later) 
that the {\it naive LF restriction} of the NG-boson field
leads to vanishing of both the NG-boson emission vertex
and the corresponding current vertex; namely, {\it the NG phase
is not realized in the LF quantization} \cite{KTY}.
 
Based on the LSZ reduction formula, 
the NG-boson emission vertex $A \rightarrow B + \pi$ may be written as 
\begin{eqnarray}
\lefteqn{\langle B \pi(q)\vert A \rangle = i\int d^4 x e^{iqx} \langle B 
\vert \Box \pi(x) \vert A \rangle} \nonumber \\
&=&i(2\pi)^4 \delta(p_A^{-}-p_B^{-}-q^-)
\delta^{(3)}(\vec{p}_A-\vec{p}_B-\vec{q})
\langle B \vert j_{\pi}(0)\vert A 
\rangle ,   
\end{eqnarray}
where $\pi(x)$ and $j_{\pi}(x)=\Box \pi(x)=(2\partial_{+}\partial_{-}
-\partial_{\bot}^2) \pi(x)$ are 
the interpolating field and the source function of the NG boson, 
respectively, and $q^{\mu}=p^{\mu}_A-p^{\mu}_B$ 
are the NG-boson four-momenta and 
$\vec{q}\equiv (q^{+},q^{\bot})$.
It is customary \cite{Wei} to take the collinear momentum frame,
 $\vec{q}=0$ and $q^{-}\ne 0$ (not a soft momentum),
for the emission 
vertex of the exactly massless
NG boson with $q^2=0$.

Then the NG-boson emission vertex should vanish on the LF
due to the periodic boundary condition:
\begin{eqnarray}
\lefteqn{(2\pi)^3\delta^{(3)}(\vec{p}_A-\vec{p}_B)
\langle B\vert j_{\pi}(0)\vert
A\rangle}\nonumber\\
&=&\int d^{2}x^{\bot}\lim_{L\rightarrow\infty}\langle B\vert 
\Bigl(\int^{L}_{-L}dx^{-}2\partial_{-}\partial_{+}\pi\Bigr)\vert A
\rangle=0 .
\end{eqnarray}

Another symptom of this disease is the vanishing of the current vertex
(analogue of $g_A$ in the nucleon matrix element).   
When the continuous symmetry is
spontaneously broken, the NG theorem requires that the corresponding 
current $J_{\mu}$ contains  
an interpolating field of the NG boson $\pi(x)$, that is,    
 $J_{\mu}=-f_{\pi}\partial_{\mu}\pi+\widehat
J_{\mu}$, where $f_{\pi}$ is the ``decay constant'' of the NG boson
and $\widehat J_{\mu}$ denotes the non-pole term.
Then the current conservation $\partial_{\mu} J^{\mu}=0$ leads to
\begin{eqnarray} 
\lefteqn{0=
\langle B \vert 
\int d^3\vec{x}\, \partial_{\mu}\widehat J^{\mu}(x)\vert A
\rangle_{x^+=0}}\nonumber\\
&=&-i(2\pi)^3\delta^{(3)}(\vec{q})\, \displaystyle{
\frac{m_{A}^2-m_{B}^2}{2p_A^+}}\langle B \vert \widehat J^+(0)
\vert A \rangle ,
\label{c-vertex} 
\end{eqnarray} 
where $\int d^3\vec{x} \equiv \lim_{L\rightarrow \infty}
\int_{-L}^{L} dx^-d^2 x^{\bot}$ and the integral of 
the NG-boson sector $\Box\pi$ has no contribution on the LF
because of the periodic boundary condition as we mentioned before. 
Thus the current vertex $\langle B\vert\widehat J^{+}(0)\vert A\rangle$
should vanish at $q^2=0$ as far as $m^2_{A}\ne m^2_{B}$.

This is actually a manifestation of the conservation of a charge
$\widehat Q\equiv
\int d^3\vec{x}\, \widehat J^{+}$
which contains only the non-pole term.
Note that $\widehat Q$ coincides with the full LF charge
$Q\equiv
\int d^3 \vec{x}\,  J^{+}$, since   
the pole part always drops out of 
$Q$ due to the integration on the LF, i.e., $Q=\widehat Q$.
Therefore the {\it conservation of} $\widehat{Q} $ {\it inevitably follows
 from the conservation of} $Q$:
$[\widehat Q, P^-]=[Q, P^-]=0$, which in fact implies vanishing 
current vertex mentioned above. This is in sharp contrast to the 
charge integrated over usual space $\mbox{\boldmath$x$}=(x^1,x^2,x^3)$ 
in the equal-time quantization: $Q^{\rm et} = 
\int d^3\mbox{\boldmath$x$} J^0$ is conserved while  
$\widehat Q^{\rm et} = \int d^3\mbox{\boldmath$x$} \widehat J^0$ is not.

Here we should emphasize that the above conclusion 
is {\it not} an artifact of DLCQ but is inherent
in the very nature of the LF quantization \cite{TY}, as far as 
we discuss the exact symmetry limit from the beginning. 

\subsection{REALIZATION OF NG PHASE}

Now,  we propose to regularize
the theory  by introducing 
explicit-breaking mass of the NG boson $m_\pi$ and then take the symmetric
limit in the end under certain condition. 
The essence of the NG phase with a small
explicit symmetry breaking can well be described by the old notion of
the PCAC hypothesis:
 $\partial_{\mu}J^{\mu}(x)=f_{\pi}m_{\pi}^2\pi(x)$, with $\pi(x)$ being
 the interpolating field of the (pseudo-) NG boson $\pi$. 
  From the PCAC relation the current 
divergence of the non-pole term $\widehat J^{\mu}(x)$ reads 
$
\partial_{\mu}\widehat J^{\mu}(x)=f_{\pi}(\Box+m_{\pi}^2)\pi(x)= 
f_{\pi}j_{\pi}(x).
$
Then we obtain 
\begin{equation}
\label{PCAC}
\langle B \vert \int d^3\vec{x}\,
\partial_{\mu}\widehat J^{\mu}(x)
\vert A \rangle 
=f_{\pi}m_{\pi}^2
\langle B \vert 
\int d^3 \vec{x}\, \pi(x)\vert A \rangle  \nonumber \\  
=\langle B \vert 
\int d^3 \vec{x}\, f_{\pi} j_{\pi}(x) 
\vert A \rangle, 
\end{equation}
where the integration of the pole term $\Box \pi(x)$ is dropped 
out as before. 
The equality between the first and the third terms
is a generalized Goldberger-Treiman relation, if both are non-zero. 
The second expression of (\ref{PCAC})
 is nothing but the matrix element of 
the LF integration of the $\pi$ zero mode (with $P^+=0$)
$\omega_{\pi} \equiv \frac{1}{2L}\int_{-L}^{L} dx^- \pi(x)$.
Suppose that $\int d^3\vec{x}\, \omega_{\pi} (x)
=\int d^3\vec{x}\, \pi (x)$ is regular when  
$m_{\pi}^2\rightarrow 0$. Then this leads to the ``no-go theorem''
 again. Thus, in order to have non-zero NG-boson emission 
 vertex (R.H.S. of (\ref{PCAC}))
  as well as non-zero current vertex (L.H.S.) at $q^2=0$, 
 the $\pi$ zero mode
$\omega_{\pi}(x)$ must behave as 
\begin{equation}
\int d^3 \vec{x}\, \omega_{\pi}\sim \frac 1{m_{\pi}^2} 
\quad (m_{\pi}^2 \rightarrow 0).
\label{omega}
\end{equation}

This situation may be clarified when the PCAC relation is written 
in the momentum space:
\begin{equation}
\displaystyle{
\frac{m_{\pi}^2f_{\pi}j_{\pi}(q^2)}{m_{\pi}^2-q^2}=
\partial^{\mu} J_{\mu}(q)=
\frac{q^2f_{\pi}j_{\pi}(q^2)}{m_{\pi}^2-q^2}
+\partial^{\mu}\widehat J_{\mu}(q). 
}
\label{PCAC-mom}
\end{equation}
What we have done when we reached the ``no-go theorem'' can be
summarized as follows:
We first set L.H.S of (\ref{PCAC-mom}) to
zero (or equivalently, assumed implicitly the regular behavior of
$\int d^3\vec{x}\, \omega_{\pi}(x)$) in the symmetric limit 
in accord with the current conservation $\partial^{\mu} J_{\mu}=0$. Then   
in the LF formalism with $\vec{q}=0 $ $(q^2=0)$, 
the first term (NG-boson pole term) of R.H.S. was also zero
due to the periodic boundary condition or the zero-mode constraint.
Thus we arrived at  $\partial^{\mu}\widehat J_{\mu}(q)=0$.
However, this procedure is
equivalent to playing a 
nonsense game: $1=\lim_{m_{\pi}^2,\,q^{2}\rightarrow 0}
(\frac{m^{2}_{\pi}-q^{2}}{
m^{2}_{\pi}-q^{2}})=0$ as far as $f_{\pi}j_{\pi}\ne 0$ (NG phase).
Therefore the {\it ``$m_{\pi}^2 =0$ theory'' with vanishing L.H.S. 
is ill-defined on the LF, namely, the ``no-go theorem'' is false}.
The correct procedure should be to 
take the symmetric limit $m_{\pi}^2 \rightarrow 0$
{\it after} the LF restriction $\vec{q}=0$ $ (q^2=0)$ 
although (\ref{PCAC-mom}) itself yields the same result 
$f_{\pi} j_{\pi} = \partial^{\mu} \widehat J_{\mu}$,
irrespectively of the order of the two limits $q^{2}\rightarrow 0$ and
$m_{\pi}^2\rightarrow 0$. Then  (\ref{omega}) does follow.

This implies that {\it at quantum level} the LF charge $Q=\widehat Q$ 
 is {\it not conserved}, or the {\it current conservation does not hold}
for a particular Fourier component with $\vec{q}=0$ even 
in the symmetric limit: 
\begin{equation}
\dot{Q}=\frac{1}{i}[Q, P^{-}]=\partial^{\mu} J_{\mu}\vert_{\vec{q}=0}=f_{\pi} 
\lim_{m_{\pi}^2\rightarrow 0} m^2_{\pi}\int d^3 \vec{x}\, \omega_{\pi} \neq 0.
\label{nonconserv}
\end{equation}

\section{THE SIGMA MODEL}

Let us now demonstrate \cite{KTY} that   
 (\ref{omega}) and (\ref{nonconserv}) indeed take place {\it
 as the solution
 of the constrained zero modes} in the NG phase
 of the $O(2)$ linear $\sigma$ model:  
\begin{equation}\label{lag}
{\cal L}=\frac{1}{2}(\partial_{\mu}\sigma)^2+\frac{1}{2}
(\partial_{\mu}\pi)^2-\frac{1}{2}\mu^2 (\sigma^2+\pi^2)-\frac{\lambda}{4}
(\sigma^2+\pi^2)^2 +c\sigma, 
\end{equation}
where the last term is the explicit breaking which regularizes the 
NG-boson zero mode. 
   
In the DLCQ we can clearly separate
the zero modes (with $P^{+}=0$), $\pi_0
\equiv \frac{1}{2L}\int_{-L}^{L} dx^- \pi(x)$ (similarly for $\sigma_0$),
from other oscillating modes (with $P^{+} \ne 0$),
$\varphi_{\pi}\equiv \pi-\pi_0$ (similarly for $\varphi_{\sigma}$).
Through the Dirac quantization of the constrained system
the canonical commutation relation for the 
oscillating modes reads \cite{MY} 
\begin{equation}
\left[\varphi_i(x),\varphi_j(y)\right]
=-{i \over 4}\{\epsilon(x^--y^-)-{x^--y^- 
\over L}\}\delta_{ij}\delta^{(2)} (x^{\bot}-y^{\bot}), 
\label{commutator}
\end{equation}
where each index stands for $\pi$ or $\sigma$. 
Comparing (\ref{commutator}) with (\ref{FSf}),
we can see that the second term in $\{ \}$ corresponds to subtracting the
zero mode contribution out of the commutator. 
 
On the other hand, the zero modes 
 are  not independent degrees of freedom but are 
implicitly determined by $\varphi_{\sigma}$ and $\varphi_{\pi}$
through the second class constraints, the
zero-mode constraints \cite{MY}:
\begin{equation}
\chi_{\pi}\equiv
\displaystyle{\frac 1{2L} \int^L_{-L}dx^-}\left[(\mu^2-\partial_{\bot}^2)\pi
+\lambda \pi(\pi^2+\sigma^2)\right]= 0,
\label{pizero}
\end{equation}
and similarly,  $\chi_{\sigma}\equiv 
\frac{1}{2L}\int^L_{-L}dx^-\{[\pi \leftrightarrow \sigma]-c\} = 0$.
Thus the zero modes are solved away from the physical Fock space which is 
constructed upon the trivial vacuum.
Note that through the equation of motion  
these constraints are equivalent  to 
the characteristic of the DLCQ with periodic boundary condition \cite{MR}:
$
\chi_{\pi}=-\frac{1}{2L}\int^L_{-L}dx^-\,2\partial_{+}\partial_{-}\pi
=0,
$ (similarly for $\sigma$)
which we have used to prove the ``no-go theorem'' for the case of
$m_{\pi}^2\equiv 0$. Thus
the `no-go'' theorem is a consequence of the
zero-mode constraint itself in the case of $m_{\pi}^2\equiv 0$. Namely,
{\it solving the zero-mode constraint does not give rise to SSB at all in
the exact symmetric case} $m_{\pi}^2\equiv 0$, in contradiction to
the naive expectation \cite{Rob}. 

Actually, in the NG phase $(\mu^2 < 0)$  the equation of motion 
of $\pi$ reads
$(\Box+m_{\pi}^2)\pi(x)=-\lambda({\pi}^3+\pi
{\sigma}'^2+2v\pi{\sigma}') $ $\equiv j_{\pi}(x)
$, with ${\sigma}'=
\sigma-v$ and $m_{\pi}^2=\mu^2+\lambda v^2=c/v$, 
where $v\equiv \langle \sigma\rangle$ is the classical 
vacuum solution  determined by  $\mu^2 v
+\lambda v^3 =c$. 
Integrating the above equation of motion over $\vec{x}$, we have
\begin{equation}
\int d^{3} \vec{x}\, j_{\pi}(x) -m_{\pi}^2 \int d^{3} \vec{x}\, 
\omega_{\pi}(x) 
=\int d^{3} \vec{x}\, \Box\pi(x)
=-\int d^{3} \vec{x}\, \chi_{\pi}=0,
\label{eqmot}
\end{equation} 
where $\int d^{3} \vec{x}\, \omega_{\pi}(x) =
\int d^{3} \vec{x}\, \pi(x)$.
Were it not for the singular behavior (\ref{omega}) for the $\pi$ zero mode
$\omega_{\pi}$, we would have concluded 
$ (2\pi)^3\delta^{(3)}(\vec{q})\,
\langle \pi \vert j_{\pi}(0) \vert \sigma \rangle=
-\langle \pi \vert \int d^3 \vec{x}\, \chi_{\pi}
\vert \sigma \rangle=0$
in the symmetric limit  $m_{\pi}^2 \rightarrow 0$.
Namely, the NG-boson vertex at $q^2=0$ would have vanished, which is
exactly what we called ``no-go theorem''
now related to the zero-mode constraint $\chi_{\pi}$.
On the contrary, direct evaluation of the matrix element of
$j_{\pi}=-\lambda({\pi}^3+\pi
{\sigma}'^2+2v\pi{\sigma}') $ 
in the lowest order perturbation yields non-zero result 
even in the symmetric limit $m_{\pi}^2\rightarrow 0$:
$\langle \pi \vert j_{\pi}(0) \vert \sigma \rangle 
=-2\lambda v \langle \pi \vert \varphi_{\sigma} \varphi_{\pi}\vert 
\sigma\rangle =-2 \lambda v \ne 0 \quad(\vec{q}=0),
$
which is in agreement with the usual equal-time formulation.
Thus we have seen that  
naive use of the zero-mode constraints by setting $m_{\pi}^2\equiv 0$ 
leads to the {\it internal inconsistency} in the NG phase.
The ``no-go theorem'' is again false.  

The same conclusion can be obtained more systematically by solving 
the zero-mode constraints in the perturbation 
around the classical (tree level) solution to the zero-mode constraints
which is nothing but the minimum of the classical potential:
$v_{\pi}=0$ and $v_{\sigma}\equiv v$, where we have
divided the zero modes 
$\pi_{0}$ (or $\sigma_{0}$) into classical constant piece $v_{\pi}$
(or $v_{\sigma}$) and operator part $\omega_{\pi}$ (or 
$\omega_{\sigma}$). 
The operator zero modes 
are solved perturbatively by substituting the expansion 
$\omega_i=\sum_{k=1}\lambda^k \omega_i^{(k)}$ into $\chi_{\pi}$,
$\chi_{\sigma}$  under the Weyl ordering.

 The lowest order solution of the zero-mode constraints $\chi_{\pi}$
 and $\chi_{\sigma}$ for $\omega_{\pi}$ takes the form:
\begin{equation}
(-m_{\pi}^2+\partial_{\bot}^2)\, \omega_{\pi}
=\frac{\lambda}{2L}\int_{-L}^{L}dx^-(\varphi_{\pi}^3
+\varphi_{\pi}\varphi_{\sigma}^2+2v\varphi_{\pi}\varphi_{\sigma}),
\label{operatorzero}
\end{equation}
which in fact yields (\ref{omega}) as
\begin{equation}
\lim_{m_{\pi}^2\rightarrow 0} m_{\pi}^2\int d^3 \vec{x}\, \omega_{\pi}
=-\lambda\int d^3 \vec{x}\, (\varphi_{\pi}^3
+\varphi_{\pi}\varphi_{\sigma}^2+2v\varphi_{\pi}\varphi_{\sigma})
\ne 0.
\label{omega2}
\end{equation}
This actually ensures non-zero 
 $\sigma \rightarrow \pi \pi$ vertex through (\ref{eqmot}): 
$ 
\langle \pi \vert j_{\pi}(0) \vert \sigma \rangle
=-2\lambda v, 
$
which agrees with the previous direct evaluation as it should.

Let us next discuss the LF charge operator corresponding to the current
$J_{\mu}=\partial_{\mu}
\sigma \pi-\partial_{\mu}\pi \sigma$.  
The LF charge  
$
Q=\widehat Q=
\int d^3\vec{x}\,
(\partial_{-}\varphi_{\sigma}\varphi_{\pi}-\partial_{-}
\varphi_{\pi}\varphi_{\sigma})
$
{\it contains no zero modes} and hence no $\pi$-pole term which was
dropped by the integration due to 
the periodic boundary condition and the $\partial_{-}$, 
so that $Q$ is well defined even in the NG phase and hence 
annihilates the vacuum simply by the $P^+$ conservation \cite{MY}:
\begin{equation}
Q \vert  0 \rangle=0.
\label{vac-annih}
\end{equation}    
This is also consistent with explicit computation of the
commutators: 
$\langle [Q, \varphi_{\sigma}]\rangle =-i \langle\varphi_{\pi}\rangle=0$ and 
$\langle [Q,\varphi_{\pi}]\rangle =i\langle 
\varphi_{\sigma}\rangle=0$ \footnote{
By explicit calculation with a careful treatment of the zero-modes
contribution we can also show that
$\langle [Q, \sigma]\rangle =
\langle [Q,\pi]\rangle =0$ \cite{TY}.
}
,
 which are contrasted to 
 (\ref{nonannihilation}) in the continuum theory. They are also to be compared
 with  those in the usual equal-time case 
 where the SSB charge does not 
annihilate the vacuum $Q^{\rm et} \vert 0\rangle\ne 0$: 
$\langle [Q^{\rm et},\sigma]\rangle
=-i\langle \pi\rangle=0,
\langle [Q^{\rm et},\pi]\rangle=i\langle \sigma\rangle\ne 0$.

Since the PCAC relation is now an operator relation for
the canonical field $\pi(x)$ with $f_{\pi}=v$ in this model, 
(\ref{omega2}) 
 ensures $[Q,P^-]\ne 0$ or a non-zero 
 current vertex $\langle\pi\vert \widehat J^{+} \vert \sigma\rangle \ne 0$ 
$ (q^2=0)$ in the symmetric limit.
 Noting that $Q=\widehat Q $, we conclude that  
the regularized zero-mode
constraints indeed lead to non-conservation of the LF charge in the
symmetric limit $m_{\pi}^2\rightarrow 0$:
\begin{equation} 
\dot{Q}=\frac{1}{i}[Q, P^-]= v 
\lim_{m_{\pi}^2\rightarrow 0} m_{\pi}^2
\int d^3\vec{x}\, \omega_{\pi}\neq 0.
\end{equation}
This can also be confirmed by direct computation of $[Q, P^-]$ through the 
canonical commutator and explicit 
use of the regularized zero-mode constraints \cite{TY}.

Here we emphasize that {\it the NG theorem does not exist on the LF}. 
Instead we found
the singular behavior (\ref{omega}) which in fact {\it 
establishes existence of
the massless NG boson coupled to the current such that
$Q|0\rangle =0$ and $\dot{Q}\ne 0$}, quite analogously to
the NG theorem in the equal-time quantization which proves
existence of the massless NG boson coupled to the current such that
$Q|0\rangle\ne 0$ and $\dot{Q}=0$ (opposite to the LF case!).
Thus the singular behavior of the NG-boson 
zero mode (\ref{omega}) (or (\ref{omega2})) may be understood as a 
remnant of the Lagrangian symmetry, an analogue of the NG 
theorem in the equal-time quantization. 

\noindent
{\large Acknowledgments}\par
I would like to thank Y. Kim and S. Tsujimaru
for collaboration. This work was supported in part by a
Grant-in-Aid for Scientific Research from the
Ministry of Education, Science and Culture (No.08640365).

\end{document}